\documentclass[prb,letter,twocolumn,english, showpacs, showkeys, floatfix]{revtex4}
\usepackage[T1]{fontenc}
\usepackage[latin9]{inputenc}
\usepackage{babel}
\usepackage{times}
\usepackage{mathptmx}
\usepackage{graphicx}% Include figure files
\usepackage{dcolumn}% Align table columns on decimal point
\usepackage{bm}% bold math

\newcommand{\Phos}{$^{31}$P }

\newcommand{\sio}{SiO$_{2}$ }

\newcommand{\Pb}{$\mathrm{P}_{\mathrm{b}}$ }

\bibpunct{[}{]}{,}{n}{}{}
\begin{document}
\title{$T_1$- and $T_2$-spin relaxation time limitations of phosphorous donor electrons
\\ near crystalline silicon to silicon dioxide interface defects}
\author{S.-Y. Paik, S.-Y. Lee, W. J. Baker}
\author{D. R. McCamey}~\email{dane.mccamey@physics.utah.edu}
\author{C. Boehme}~\email{boehme@physics.utah.edu}
\affiliation{University of Utah, Department of Physics and
Astronomy, 115 South 1400 East, Salt Lake City, Utah}
\date{\today}
%%%%%%%%%%%%%%%%%%%%%%%%%%%%%%%%%%%%%%%%%%%%%%%%%%%%%%%%%%%%%%%%%%%%%
%ABSTRACT
%%%%%%%%%%%%%%%%%%%%%%%%%%%%%%%%%%%%%%%%%%%%%%%%%%%%%%%%%%%%%%%%%%%%%

\begin{abstract}
A study of donor electron spins and spin--dependent electronic
transitions involving phosphorous ($^{31}$P) atoms in proximity of
the (111) oriented crystalline silicon (c-Si) to silicon dioxide
(SiO$_{2}$) interface is presented for [$^{31}$P] = 10$^{15}$
$\mathrm{ cm}^{-3}$ and [$^{31}$P] = 10$^{16}$ $\mathrm{ cm}^{-3}$
at about liquid $^4$He temperatures ($T = 5$ $\mathrm{K} - 15$
$\mathrm{K}$). Using pulsed electrically detected magnetic
resonance (pEDMR), spin--dependent transitions between the \Phos
donor state and two distinguishable interface states are observed,
namely (i) \Pb centers which can be identified by their
characteristic anisotropy and (ii) a more isotropic center which
is attributed to E$^\prime$ defects of the \sio bulk close to the
interface. Correlation measurements of the dynamics of
spin--dependent recombination confirm that previously proposed
transitions between \Phos and the interface defects take place.
The influence of these electronic near--interface transitions on
the \Phos donor spin coherence time $T_2$ as well as the donor
spin--lattice relaxation time $T_1$ is then investigated by
comparison of spin Hahn--echo decay measurements obtained from
conventional bulk sensitive pulsed electron paramagnetic resonance
and surface sensitive pEDMR, as well as surface sensitive
electrically detected inversion recovery experiments. The
measurements reveal that both $T_2$ and $T_1$ of \Phos donor
electrons spins in proximity of energetically lower interface
states at $T\leq 13$ K are reduced by several orders of magnitude.
\end{abstract}
%71.55.Cn    Electronic structure of bulk materials: Impurity and defect levels: Elemental semiconductors

%72.25.Rb    Electronic transport in condensed matter: Spin polarized transport: Spin relaxation and scattering

%73.20.Hb    Electronic structure and electrical properties of
%surfaces, interfaces, thin films, and low-dimensional structures:
%Electron states at surfaces and interfaces: Impurity and defect
%levels; energy states of adsorbed species

%76.30.Da Magnetic resonances and relaxations in condensed matter,
%Mössbauer effect: Electron paramagnetic resonance and relaxation:
%Ions and impurities: general


 \maketitle
%%%%%%%%%%%%%%%%%%%%%%%%%%%%%%%%%%%%%%%%%%%%%%%%%%%%%%%%%%%%%%%%%%%%%
\section{Introduction}
%%%%%%%%%%%%%%%%%%%%%%%%%%%%%%%%%%%%%%%%%%%%%%%%%%%%%%%%%%%%%%%%%%%%%
Due to their technological importance, the properties of
phosphorous ($^{31}$P) donors in crystalline silicon have been
investigated extensively for more than half a century. During this
time, magnetic resonance based methods have revealed many aspects
of the microscopic nature of \Phos as well as the electronic
processes in which it is involved. Since the first mapping of a
donor wavefunction using an electron nuclear double resonance
technique \cite{feher1956}, a large and diverse amount of
information regarding the electronic and quantum mechanical
properties of such systems has been reported \cite{Feher1959}.
This wealth of information has contributed to the technological
exploitation of silicon to the extent that it has become the most
widely utilized semiconductor in the global electronics industry.
However, even with all the information regarding silicon available
to us, there are still questions regarding the ability to exploit
the quantum mechanical nature (specifically spin) of dopants and
charge carriers for technological applications such as spin
transistors~\cite{Appelbaum2007} or quantum computers~\cite{kane},
the latter of which are proposed to utilize electron or nuclear
spins of phosphorus donors in silicon as quantum bits. While there
has been significant experimental effort for the implementation of
these and other concepts~\cite{Clark}, challenges remain in a
number of different areas, the most pressing being the difficulty
in detecting the spin of individual donors without reducing its
quantum memory time (the coherence time, $T_2$)~\cite{Huebl2008}.
Nevertheless, donors in silicon retain promise in this area due to
their extremely long phase coherence times, with $T_2$ exceeding
60 ms for the electron spin~\cite{Tyryshkin2003} and 1 second for
the nuclear spin~\cite{Morton2008} at liquid He temperatures.

Many of the recent schemes for silicon based spin- or quantum
devices involve electronic processes occurring at or near
interfaces, particularly the c-Si/\sio interface. This presents
advantages since locating spins near interfaces allows them to be
controlled with surface gates~\cite{kane}, and detected with
surface
electronics~\cite{Boehme2002,stegner06,Schenkel2006,Chan2006,Buehler2006}.
However, it may also lead to a decrease in spin coherence due to
the spin--spin interactions with surface states
\cite{Schenkel2005,Sousa2007}, as well as to the loss of quantum
information following spin--dependent recombination through
surface states. Since it is not possible to obtain defect free
c-Si/\sio interfaces~\cite{lucovsky99} (we note that work is
currently being undertaken to obtain defect free interfaces using
H-termination but it is in the very early
stages\cite{Eng2005,Eng2007}) and since there are even proposals
to use interface defects as probe spins to
readout~\cite{Boehme2002} the spin of a single donor, the
understanding of interface electron and spin transitions has
become important.

In the following, pulsed electrically detected magnetic resonance
(pEDMR) and pulsed electron paramagnetic resonance (pEPR)
measurements performed on (111) surface oriented c-Si samples with
[P]=10$^{15}$ $\mathrm{cm}^{-3}$ and [P]=10$^{16}$
$\mathrm{cm}^{-3}$ at temperatures between $T = 5$ K and $T = 15$ K
are presented. The data was collected in order to elucidate the
nature of spin--dependent electronic interface transitions involving
\Phos donor and interface states and to then determine how these
processes influence the coherence time of the \Phos donor electron
spins in proximity of these interface defects. The latter was
accomplished by comparison of the interface sensitive pEDMR
measurements to bulk sensitive pEPR measurements of \Phos donors.
Our results are discussed with regard to their implications for the
ability of spin qubit readout using interface defect probe spins
whilst maintaining long coherence times. We emphasized that while
the key questions motivating this study are centered about \Phos
qubit coherence times, the study presented follows an extensive
number of previous
EPR~\cite{Nish:1971,Poin:1981,poin:1984,stes:1987,Kim:1988,stes:1991,lena:1998},
EDMR~\cite{lena:1998,Kan3,stegner06,McCamey2006,mcc:prb:2008} and
pEDMR~\cite{Frie:05,Boe:TSF,Boe05} studies carried out on various
c-Si/\sio interface defects as well as electronic trapping and
recombination processes of interfaces with different surface
orientations and \Phos doping concentrations. Most of these studies
aimed to enhance the understanding of electronic processes relevant
for materials systems used in conventional c-Si based
microelectronics and photovoltaic devices. Thus, the study presented
here may also be of relevance for conventional silicon technologies.

%%%%%%%%%%%%%%%%%%%%%%%%%%%%%%%%%%%%%%%%%%%%%%%%%%%%%%%%%%%%%%%%%%%%%
\section{Experiments}
\begin{figure}[tl]
\includegraphics[width=\columnwidth]{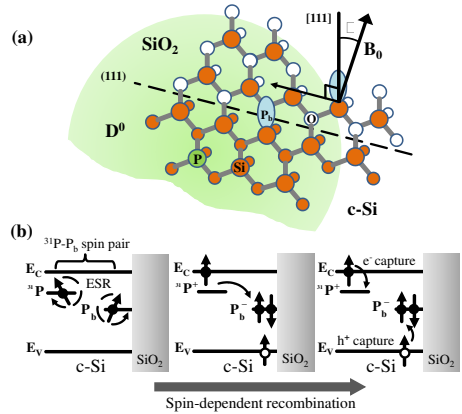}
\caption{(color online) (a) Ball and stick illustration of the
atom scale structure of the c-Si/\sio interface. The \Phos donor
state (green circle) is significantly larger than the \Pb center.
Exchange coupling between the two states is possible when they are
sufficiently close to each other. The orientation of the interface
is defined by the angle $\theta$ between the externally applied
$B_0$ magnetic field and the (111) crystal axis which is
perpendicular to the crystal surface. (b) Band sketch illustrating
the spin--dependent \Phos/\Pb charge carrier recombination
transition. For details see text} \label{fig:1}
\end{figure}

%%%%%%%%%%%%%%%%%%%%%%%%%%%%%%%%%%%%%%%%%%%%%%%%%%%%%%%%%%%%%%%%%%%%%
Following the recent demonstration of electrical detection of
\Phos spin states~\cite{stegner06} we anticipate, for the pEDMR
measurements presented here, the presence of spin--dependent
recombination between \Phos donor states and energetically lower
c-Si/\sio interface states, as illustrated for the example of the
\Pb interface state in Fig.~\ref{fig:1}. Figure~\ref{fig:1}(a)
displays a geometrical stick and ball sketch of the c-Si/\sio
interface where some of the silicon (111) surface atoms posses
unsaturated bonds (called dangling bonds). These dangling bonds
are all oriented along the (111) axis~\cite{lena:1998} and are
highly localized states with much p-content and only small
s-content~\cite{brow:1983,cook:1988}. Silicon dangling bonds
oriented along the (111) direction (also called \Pb centers) are
highly anisotropic and, since they are paramagnetic when
uncharged~\cite{Nish:1971,Nish:1972}, can be identified using EPR
by their well known Land\'e (g) factor
anisotropy~\cite{lena:1998}, revealed by measurement of EPR
spectra as a function of the angle $\theta$ between the $B_0$
field and the (111) direction (see sketch). In contrast to the \Pb
state, the shallow \Phos donor state (illustrated by the large
shaded circle in Fig.~\ref{fig:1}(a)) is a significantly less
localized state having an s-like envelope function with a
localization length of $\approx 3\mathrm{nm}$, strongly modulated
by the periodic crystalline structure of its host
environment~\cite{Koiller,Wellard2005}. Since the \Phos donor
state encompasses thousands of silicon sites, an exchange
interaction between near--surface \Phos donors and \Pb centers can
occur allowing the transition of the donor electron into the
energetically lower interface state. This transition is sketched
for an interface band diagram in Fig.~\ref{fig:1}(b). The
illustration shows that: (i) Both the \Phos donor state as well as
the \Pb state become charged (positively and negatively,
respectively) through the transition. (ii) Because of the Pauli
principle, and spin conservation due to the weak spin--orbital
coupling of silicon, the transition is spin--dependent. Hence, the
transition provides a spin--to--charge conversion mechanism and as
a result, since the charge of the \Phos$^+$/\Pb$^-$ can be
detected through the measurement of recombination currents,
\Phos/\Pb pairs have been proposed as electric readout mechanism
for \Phos qubits~\cite{Boehme2002,stegner06}.

%|figure 1

Experimentally, time domain measurements of spin-dependent
c-Si/\sio--interface recombination were conducted in order to
verify that the qualitative behavior of interface currents after
EPR excitation of the \Phos or the \Pb states is similar to higher
doped c-Si samples with (100) orientation. In order to also verify
electrically detect spin--Rabi oscillation (as expected from
previous studies~\cite{stegner06,Huebl2008}),
transient-nutation-like measurements of the integrated current
transients as a function of the excitation lengths were carried
out. Following these preparatory measurements, four experiments
were performed to address the following questions:

(i) Is the EDMR signal that has previously been
observed~\cite{stegner06} at magnetic fields between the two
hyperfine split lines of the \Phos for measurements conducted at
X-band excitations (approx. 10 GHz) truly due to \Pb states? If so,
are \Pb states the only interface states involved in spin--dependent
transitions as illustrated in Fig.~\ref{fig:1}(b) or are there other
states at or near the interface, or even in the bulk, which could
contribute to the observed signals? In order to address this
question, a systematic study of the EDMR spectrum (the magnetic
field dependence) as a function of the interface orientation angle
$\theta$ was made in order to observe all $g$--factors involved in
spin--dependent recombination and to detect possible anisotropies of
these centers.

(ii) Are the observed spin--dependent processes truly due to pairs
of \Phos and interface states as depicted in Fig.~\ref{fig:1}(b)
or are the signals corresponding to different $g$--factors due to
independent processes? In order to address this question, a series
of different spin--dependent current transients was recorded under
various temperatures, light intensities, surface orientations,
sample voltages and offset currents as well as excitation powers
and lengths. For all applied conditions, the current transients
were recorded after (a) resonant excitation of the two \Phos lines
and (b) the interface states and the correlation between the
dynamics of \Phos and interface states were compared.

(iii) As an understanding about the nature of the interface
defects and the transitions between interface defects and the
\Phos donor electrons was established, the main question of this
study was addressed: Are the coherence times of \Phos donors near
interface defects compromised? For this, a series of coherence
time ($T_2$) measurements was carried out on \Phos donors in
proximity to interface states and with the same samples, under the
same conditions (in fact, during the same experimental runs), the
coherence time of bulk \Phos was measured for comparison. For the
interface $T_2$ measurements, modified Hahn echoes, detected with
pEDMR, were used, in a similar way as recent studies of \Phos
doped c-Si samples with (100) orientation~\cite{Huebl2008} and
spin--dependent \Phos bulk processes at very high magnetic
fields~\cite{Morley2008a}. For the bulk $T_2$ measurements,
conventional ESR detected Hahn--echo experiments were carried out.
This comparative study of $T_2$ times was made as a function of
the temperature for $5$ $\mathrm{K}\leq T\leq 13$ $\mathrm{K}$.

(iv) Finally, in order to obtain information on wether the
spin--coherence times $T_2$ of near--interface defect \Phos is
determined by the electronic transitions between \Phos and
interface states or by the interface defect induced spin--spin
relaxation processes as suggest by de Sousa~\cite{Sousa2007}, a
comparison of electrically detected $T_2$ times and $T_1$ times
was made for a temperature of $T=5\mathrm{L}$. In order to measure
$T_1$, electrically detected inversion recovery measurements were
performed. To the knowledge of the authors, this is the first time
that such an experiment is demonstrated.

%%%%%%%%%%%%%%%%%%%%%%%%%%%%%%%%%%%%%%%%%%%%%%%%%%%%%%%%%%%%%%%%%%%%%
\section{Experimental details}
%%%%%%%%%%%%%%%%%%%%%%%%%%%%%%%%%%%%%%%%%%%%%%%%%%%%%%%%%%%%%%%%%%%%%
For the experiments presented in the following, we used 300 $\mu$m
thick, (111) surface oriented silicon since \Pb centers for this
surface orientation are all identically oriented. This is in
contrast to the (100) surface previously used for similar
experiments~\cite{stegner06,Huebl2008} as the (100) surface allows
silicon dangling bonds (here they are called $\mathrm{P_{b0}}$
states) to exist in two orientations at the same time which makes
their mutual EPR spectroscopic distinction as well as their
distinction from other possible interface defects very difficult.
The experiments were conducted with dopant concentrations of [P]
$=10^{15}$ $\mathrm{ cm}^{-3}$ and $10^{16}$ $\mathrm{ cm}^{-3}$
which is less than $10^{17}$ $\mathrm{ cm}^{-3}$ of previous EDMR
studies~\cite{stegner06,Huebl2008}. The lower concentrations ensured
that interactions between neighboring \Phos atoms which are known to
exist~\cite{FEHER1959b} at [P] $=10^{17}$ $\mathrm{ cm}^{-3}$ can be
neglected.

In order to enable the electrical detection of spin--coherence,
the c-Si samples required contact structures which would not
strongly distort the homogeneity of the spin resonant microwave
fields $B_1$. This problem was solved using a method similar to previous
pEDMR studies at X-band~\cite{stegner06,mcc:nmat:2008} - we
designed long match--like sample substrates on which electrical
sample contacts outside of the microwave field are connected to
the sample via $\sim50$ mm long thin--film Al wires whose
thickness of 100 nm is below the penetration depth of the applied
microwave radiation. The contact structures were fabricated using
a photo lithographical lift--off process that was carried out
after the Al film was evaporated on the H-terminated silicon (111)
surface of the c-Si samples that had been prepared by a wet
treatment with hydrofluoric acid. In order to maximize the density
of interface states, a native oxide was grown on the (111) surface
after the contact deposition by exposure of the sample to ambient
air.

All EDMR and EPR experiments were carried out at X-band using a
cylindrical dielectric low--Q pulse resonator which was part of a
Bruker Elexsys E580 EPR spectrometer. The sample temperatures were
obtained with a $^4$He flow cryostat, the excess charge carriers
were induced through a spectral cold light source (IR and UV
filtered spectral light) with an incandescent light source (Schott
KL 2500 LCD) with lamp temperature of about $3000$ K producing an
integrated spectral intensity of approximately $5$ $\mathrm{ W
cm^{-2}}$ at the sample surface. The EDMR experiments were conducted
by establishing a constant offset photocurrent using a constant
current source with a time constant in excess of the experimental
shot repetition time. Current transients were digitized and recorded
following current amplification by a Stanford Research SR570.

The raw data recorded for the presented measurements was a
combination of spin--dependent currents and microwave induced
artifact currents. The latter can be recorded separately by
measurement of the current response at off--spin resonant
$B_0$--fields. Magnetoresistance effects on the microwave induced
currents can be linearly extrapolated for c-Si at the given
magnetic fields. The microwave current transients obtained from
this procedure were subtracted from the raw data in order to
reveal the current transients solely caused by
spin--dependent transitions.

%%%%%%%%%%%%%%%%%%%%%%%%%%%%%%%%%%%%%%%%%%%%%%%%%%%%%%%%%%%%%%%%%%%%%
\section{Measurement Results}
%%%%%%%%%%%%%%%%%%%%%%%%%%%%%%%%%%%%%%%%%%%%%%%%%%%%%%%%%%%%%%%%%%%%%

%%%%%%%%%%%%%%%%%%%%%%%%%%%%%%%%%%%%%%%%%%%%%%%%%%%%%%%%%%%%%%%%%%%%%
\subsection{Identification of spin--dependent transitions}
%%%%%%%%%%%%%%%%%%%%%%%%%%%%%%%%%%%%%%%%%%%%%%%%%%%%%%%%%%%%%%%%%%%%%
\label{idspdeptrans}
\subsubsection{Experimental data}
In order to confirm the results of Stegner et al.~\cite{stegner06}
for a c-Si:P/\sio interface with smaller \Phos concentration and a
(111) surface orientation of the silicon sample, transient
measurements of photocurrent changes $\Delta I$ were recorded under
various illumination conditions and temperatures. Figure~\ref{fig:2}
displays a data set of $\Delta I(B_0,t)$ recorded as a function of
the magnetic field $B_0$ and the time $t$ after a 96 ns long
microwave pulse with a frequency of $f=9.749$ GHz and a power of
$P\approx 8$ $W$ for $T=5$ K and a constant photocurrent of $I=270$
$\mu\mathrm{A}$. The sample orientation was $\theta=0$. The data set
clearly confirms the expected EPR induced currents with three local
response maxima at $B_0=346.37$ mT, $B_0=347.9$ mT and $B_0=350.55$
mT. The two outer peaks which are separated by a magnetic field of
$\approx 4.2$ mT are the two hyperfine lines of the \Phos donor
electron while the peak close to the low field \Phos line has been
attributed to interface defect states~\cite{stegner06}.

%%%%%%%%%%%%%%%%%%%%%%%%%%%%%%%%%%%%%%%%%%%%%%%%%%%%%%%%%%%%%%%%%%%%%
\subsubsection{Discussion}
\label{maindist} 
\begin{figure}[b]
\centering\includegraphics[width=\columnwidth]{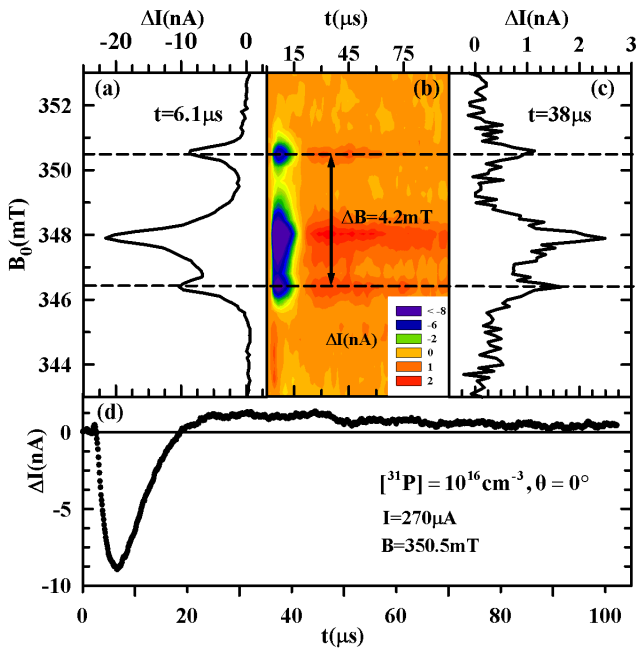}
\caption{(color online) Plots of the microwave pulse (at $t=0$)
induced change $\Delta I$ of an interface photocurrent along a
c-Si(111):P/\sio interface as a function of time $t$ and the
applied magnetic field $B_0$. (a,c) Plots of $\Delta I$ as
functions of $B_0$ for the times $t=6.1$ $\mu$s and $t=38$ $\mu$s,
respectively, for which a quenching and an enhancement of the
current changes reach their extrema. (b) color contour plot of the
entire data set $\Delta I(t,B_0)$ containing the data of the plots
in (a), (c) and (d). (d) Plot of $\Delta I$ as a function of the
time $t$ for a magnetic field $B_0=350.5$ mT which was on
resonance with the high field \Phos EPR line.} \label{fig:2}
\end{figure}

The transient behavior at the magnetic fields with
EPR responses confirms the measurements by Stegner et al.
\cite{stegner06}. It consists of a brief photocurrent quenching
after the pulse, attributed to an enhancement of the interface
recombination, followed by a longer-lived current enhancement. This
enhancement arises because the singlet recombination rate returns to
its steady state faster than the resonantly quenched triplet
recombination rate, causing a net quenching of the recombination
rate~\cite{stegner06,Boe03b}. Note that variations of the
temperature, the sample voltage and therefore the photocurrent as
well as the illumination conditions change the quantitative dynamics
of the observed transients - they do not, however, change the
qualitative quenching/ enhancement behavior displayed by the data
set of Fig.~\ref{fig:2}, which were also observed in previously
reported measurements~\cite{stegner06} on c-Si(100):P/\sio with
$\mathrm{[P]}=10^{17}\mathrm{cm^{-3}}$.

%figure 2

 Both the quenching and the enhancement shown in
Fig.~\ref{fig:2}(d) are well fit with simple exponential decay
functions, in agreement with the observation of Stegner et al
\cite{stegner06}. This is somewhat counterintuitive since the
random spatial distribution of the paramagnetic states involved in
these transitions (see discussion in section~\ref{assign})
suggests that the distances between pairs of paramagnetic states
between which spin--dependent transitions occur are widely
distributed. This distance distribution of states also implies a
distribution of transition times~\cite{bara1} which suggests that
the observed current transients should be distributions of
exponential functions for which fits with single exponentials
would be poor. In contrast, observed transients are in good
agreement with single exponential fits which suggests that only a
narrow range of transition times exist and therefore, only pairs
with a narrow range of intra--pair distances contribute to the
observed signals. We attribute the existence of a "main--pair
distance" which dominates the observed signals to two factors:
first, pEDMR signals vanish for pairs with very large distances
(larger than the localization length of the two paramagnetic
states~\cite{bara1}) as the probability for recombiation is
greatly diminished. Second, for very short distances, where the
exchange between the two states exceeds the Larmor frequency
difference within the pair~\cite{gliesche-2008}, the signal
vanishes as the resonantly induced change of spin pair states
between singlet and triplet configurations becomes increasingly
forbidden~\cite{gliesche-2008}. Thus, there will always be a
finite main--pair distance where spin--resonance induced rate
changes become maximal and pairs around this distance will
dominate the observed spin--dependent currents.

%%%%%%%%%%%%%%%%%%%%%%%%%%%%%%%%%%%%%%%%%%%%%%%%%%%%%%%%%%%%%%%%%%%%%
\subsection{Identification of interface defects}
%%%%%%%%%%%%%%%%%%%%%%%%%%%%%%%%%%%%%%%%%%%%%%%%%%%%%%%%%%%%%%%%%%%%%
\label{assign}

\subsubsection{Experimental data}
\begin{figure}[b]
\centering\includegraphics[width=\columnwidth]{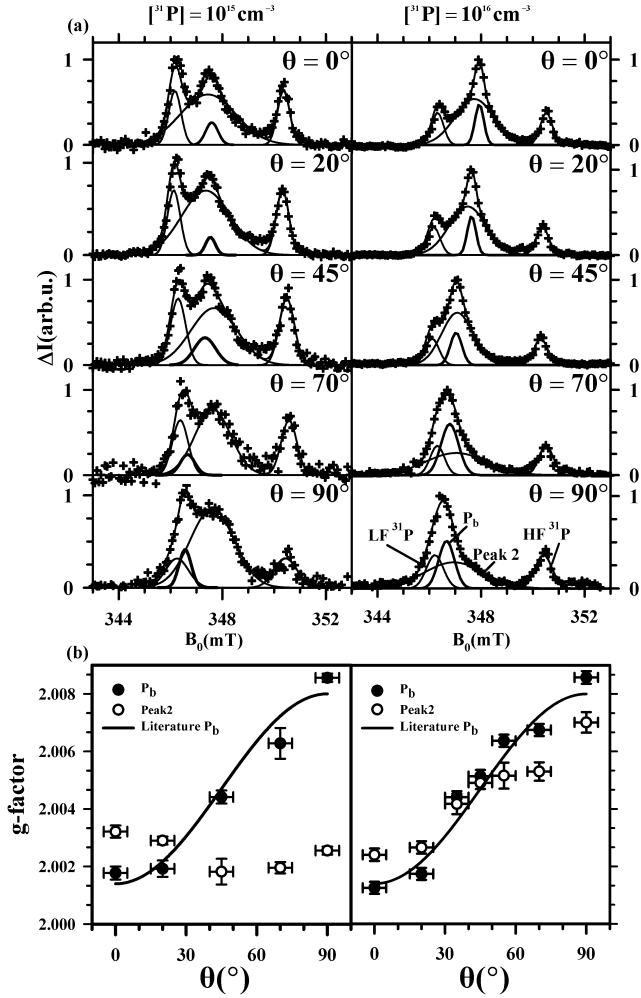}
\caption{\label{fig:4}(a) Plots of $\Delta I$ as functions of
$B_{0}$ at arbitrary times $t$ the after a microwave pulse with
arbitrary length $\tau$, frequency $f \approx 9.5$ GHz and a power
$P = 8$ W and under otherwise identical conditions as for the data
in Fig.~\ref{fig:2}. The data was collected for five sample
orientations $\theta$ and two \Phos concentrations. The displayed
plots are normalized to the maximum of $\Delta I$. The solid lines
represent fits of the data consisting of four Gaussian peaks, two
related to the \Phos hyperfine line and two peaks related to
interface defects. The plot for [\Phos]$=10^{16}\mathrm{cm}^{-3}$
and $\theta=90^o$ has the peak assignments to the low field (LF)
and high field (HF) \Phos resonances, the \Pb peak as well as peak
2. (b) Plots of the fit results of the $g$--factors for the two
interface defect lines as a function of $\theta$ for the two \Phos
concentrations. The solid lines indicate the literature values for
the \Pb center.}
\end{figure}

The data presented in Fig.~\ref{fig:2} confirms that spin--dependent
interface recombination processes can be observed with pEDMR at the
c-Si(111):P/\sio interface which are qualitatively similar to those
seen previously at the c-Si(100):P/\sio interface with higher \Phos
concentration. It is therefore possible to systematically study the
nature and the origin of the resonances found. Figure~\ref{fig:4}(a)
displays the magnetic field dependence of the pulse induced
photocurrent change $\Delta I(B_0)$ at times $t$ after the pulse
when the photocurrent changes were maximal (in the following
referred to as pEDMR spectra) for five different surface orientation
angles ($0^o\leq\theta\leq 90^o$) and two \Phos concentrations. Note
that while $\Delta I<0$, the data has positive signs as each
displayed spectrum was normalized to its respective extremum. The
normalization was made for better comparison of the data sets since
the signals obtained from samples with $\mathrm{[P]}=10^{16}$
$\mathrm{cm^{-3}}$ were significantly stronger compared to signals
from samples with $\mathrm{[P]}=10^{15}$ $\mathrm{cm^{-3}}$ (note
the higher relative noise in the latter spectra).

%figure 3

The fits of the pEDMR spectra required at least four different
Gaussian peaks. A comparison with the \Phos EPR and EDMR spectra
found in the literature~\cite{FEHER1959a,stegner06} allows
immediate identification of the peaks at the highest and lowest
magnetic fields as the well known hyperfine split \Phos donor
electron resonances. There are at least two additional resonance
signals present. As these additional peaks are very close to each
other and to the low field \Phos hyperfine resonance, significant
ambiguity for the 12 fit parameters (peak centers, width,
magnitude for all four lines) is present. This was overcome by a
stepwise fit of the spectra: we first determine the \Phos spectrum
and subsequently fit the two remaining non--phosphorous lines
(eliminating six fit parameters). The separation of the low--field
\Phos--hyperfine line from the strongly overlapping
non--phosphorous lines was achieved by first fitting the high
field \Phos hyperfine peak, which has little or no overlap with
the other resonances, with a single Gaussian line. From the result
of this fit, we can determine both the position (assuming a
hyperfine splitting of $A=4.18$ mT as verified by bulk EPR
measurements) and shape (assuming only negligible nuclear
polarization, which is justified for the given sample temperatures
and magnetic fields $B_0$) of the low--field \Phos--hyperfine
peak. The residue following these fits is then able to be fit with
two Gaussian resonances. The full fit, as well as the 4
constituent peaks, is shown for all spectra in Figure \ref{fig:4}.

Figure~\ref{fig:4}(b) shows the $g$-factors of the non-\Phos lines
as a function of the angle $\theta$, obtained from the fit. It also
displays a solid line which represents literature values for
EPR~\cite{lena:1998} and EDMR~\cite{Boe05,Frie:05} detected \Pb
centers.

%%%%%%%%%%%%%%%%%%%%%%%%%%%%%%%%%%%%%%%%%%%%%%%%%%%%%%%%%%%%%%%%%%%%%
\subsubsection{Discussion}
The anisotropy and absolute value of the $g$-factor of one of the
two peaks is  in excellent agreement with the \Pb literature values
for both \Phos concentrations. Based on this agreement, this peak
can be assigned to spin--dependent transitions (recombination) which
involve \Pb centers. Note that previous pEDMR measurements on
c-Si(111)/\sio interfaces with no \Phos doping have shown \Pb
signals~\cite{Boe05,Frie:05}. Since there are no \Phos pEDMR lines
in intrinsic c-Si, it is clear that spin--dependent \Pb interface
recombination does not necessarily require the presence of \Phos
atoms. Thus, the \Pb involvement revealed by the data in
Fig.~\ref{fig:4} may either involve \Phos as illustrated in
Fig.~\ref{fig:1}(b) or be due to an independent interface
recombination process.

In contrast to the \Pb resonance line, the assignment of the
second non-\Phos peak (referred to as peak 2, see peak assignment
in Fig.~\ref{fig:3}) is less straight forward. The fit results for
$\mathrm{[P]}=10^{15}$ $\mathrm{cm^{-3}}$ suggests that peak two
is isotropic, or only weakly anisotropic, whereas the data for
$\mathrm{[P]}=10^{16}$ $\mathrm{cm^{-3}}$ is consistent with an
anisotropic peak. There are two explanations for this difference
between the two sets of spectra, (i) that the nature of peak 2 is
different at higher \Phos concentrations, suggesting that the
observed processes may be different and (ii) that the fit error of
the center $g$-factor of peak 2 becomes increasingly inaccurate
for larger values of $\theta$. The latter may be due to the width
of peak 2 and the relative weakness of the peak 2 intensity in
comparison to the signal strengths of the \Pb and \Phos EDMR
signals at higher dopant concentrations, becoming increasingly
problematic for the fit accuracy at large $\theta$ as three
resonances (low field \Phos, \Pb, peak 2) have larger overlap at
higher angles. Note that the error ranges given in the plots of
Fig.~\ref{fig:4}(b) are based on the uncertainty estimates of the
fit routines. As additional uncertainties with regard to fit
induced systematic errors are likely to exist, the true standard
deviation for $g$--factors of peak 2 may be significantly larger.
Thus, it is not clear whether the different results for
$g$-factors of peak 2 obtained for different \Phos concentrations
are real or fit artifacts and therefore an unambiguous statement
about the anisotropy of peak 2 is not possible.

Previous pEDMR spectra on intrinsic c-Si have shown a second
isotropic recombination signal with $g\approx
2.0023(6)$~\cite{Boe05}. This is in good agreement with peak 2
observed on c-Si samples with $\mathrm{[P]}=10^{15}$
$\mathrm{cm^{-3}}$. As with the previous study on intrinsic
silicon, it is difficult to unambiguously assign peak 2 to a
particular interface defect type. Due to its strong inhomogeneity
(the large line width) and its overall magnitude in comparison to
the other pEDMR lines, it is possible that peak 2 is due to one or
more randomly oriented anisotropic centers whose powder spectra
would cause a macroscopically isotropic wide resonance line. The
$g$-factor and the line width are close (yet not equal) to the
$g$-factor and line width of silicon dangling bonds in amorphous
\sio (so called E$'$ centers~\cite{lena:1998}). Thus, it is
possible that the observed pEDMR data is due to transitions
involving E$'$ centers in proximity of the c-Si:P/\sio interface.
Note that there are several distinct E$^\prime$ centers which
differ by the backbonds of the Si atom on which the center is
located (one or more oxygen atoms, hydrogen atoms) or their
relaxation state (the unrelaxed E$'$ center has been called
$\mathrm{E}^\prime_\delta$)~\cite{lena:1998}. Charge carrier
trapping and recombination may work similarly for both the \Pb
centers with well defined orientation as well as the randomly
oriented E$'$. Due to the localization length of the \Phos donor
electron state~\cite{Koiller}, it is also conceivable that
transitions between the shallow donor states and the deep
interface states are possible. Hence, peak 2 is assigned in the
following to an E$'$ center with unspecified nature except for its
proximity to the c-Si:P/\sio interface.

%%%%%%%%%%%%%%%%%%%%%%%%%%%%%%%%%%%%%%%%%%%%%%%%%%%%%%%%%%%%%%%%%%%%%
\subsection{Identification of \Phos to interface--defect transitions}
%%%%%%%%%%%%%%%%%%%%%%%%%%%%%%%%%%%%%%%%%%%%%%%%%%%%%%%%%%%%%%%%%%%%%
\label{idphindeftrans}
\subsubsection{Experimental data}
The plots of Fig.~\ref{fig:4} demonstrate the involvement of at
least three qualitatively different paramagnetic electron states
(\Phos, \Pb, E$'$) in spin--dependent charge carrier recombination
transitions. This observation, however, does not prove that the
observed recombination transitions take place \textit{between} these
different states. While the spin--dependency of transitions through
localized states requires the existence of pairs of paramagnetic
states~\cite{Kap78}, there are examples of spin--dependent
transitions which produce only a single resonance line in EDMR
experiments~\cite{mcc:nmat:2008,mcc:prb:2008} when transitions occur
between identical centers or when the paramagnetic pairs are
strongly coupled. Therefore, the detection of multiple EDMR lines
(as in Figs.~\ref{fig:2} and ~\ref{fig:4}) leaves the exact nature
of these transitions elusive.

%figure 4
\begin{figure}[b]
\centering\includegraphics[width=\columnwidth]{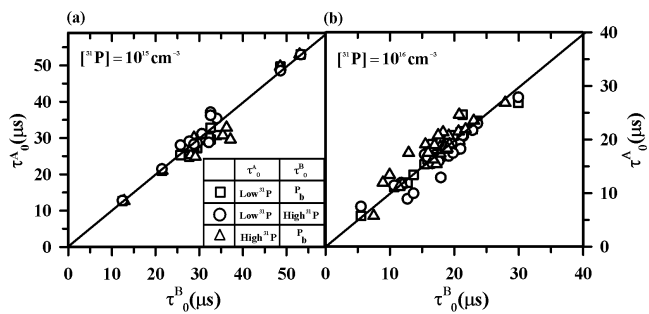}
\caption{\label{fig:5} Plots of the zero crossing times $\tau_0$
of magnetic resonantly induced photocurrent transients of the
\Phos low or high field resonance ($\tau_{0}^{A}$) versus the zero
crossing times of magnetic resonantly induced current transients
of the \Pb/E' center resonances ($\tau_{0}^{B}$) measured for a
variety of different samples and sample conditions (temperature,
offset current, illumination). The solid line is a linear function
through the origin with slope 1. The two plots made for the two
concentrations show a strong correlation of the zero crossing
times of all three resonances.}
\end{figure}

One approach to an understanding of whether the observed $g$-factors
belong to the same or different transitions is to analyze the
dynamics of the spin--dependent processes associated with these
different resonances: Spin--selection rules usually discriminate
permutation symmetries of spin $s=\frac{1}{2}$
pairs~\cite{Boe03b,gliesche-2008} which means that it is the mutual
orientation of the two spins, not the individual spin state of one
of the two pair partners, which determines the transition rate.
Therefore, the transient behavior of spin--dependent transition
rates exhibits an identical behavior after a spin resonant
manipulation of either one of the two pair partners. If two EDMR
detected resonances exhibit a different transient behavior after the
same pulsed excitation, the spin--dependent transitions
corresponding to these resonances must be different as well.
However, if the transient behavior is identical, the two resonances
may belong to identical transitions. While different dynamics of
spin--dependent transition rates for different $g$-factors is proof
that they belong to different processes, identical transition rates
only indicate that the two resonances may be due to the the same
process, as they may also be due to different processes which
coincidentally have the same transient behavior. Therefore, the
observation of identical transients requires further testing of the
correlation of the observed dynamical behavior under varying
experimental conditions.

Figure~\ref{fig:2}(b) demonstrates that the fast relaxing current
quenching and the slowly relaxing current enhancing behavior that
is discussed above occurs similarly for both the \Phos hyperfine
resonances and the resonances associated with interface states.
The dynamics (which means the decay times of the different
exponential functions), but not necessarily the absolute magnitude
of current transients from different centers involved in the same
electronic process are identical. Thus a comparison of different
transients yields information about the paramagnetic centers
between which transitions take place. We compare the
"zero--crossing time", $\tau_0$, defined as the time after the
pulsed excitation of a spin--resonance induced current transient
when the quenching and enhancement are identical. The comparison
as shown in Fig.~\ref{fig:2}(b) clearly reveals identical
$\tau_0\approx 18$ $\mu$s with an error of $\approx 1$ $\mu$s.
This suggests that the processes connected to these resonances are
due to transitions involving both the \Phos donor states as well
as the \Pb interface state or the E' near--interface state in the
way depicted in the sketch of Fig.~\ref{fig:1}(b).

In order to test whether $\tau_0$ for the interface defects and the
\Phos remained identical when the dynamics of the spin--dependent
current signal is changed, the experiment presented in
Fig.~\ref{fig:2}(b) was repeated under various combinations of
temperatures (5 K, 8 K, 10 K, 13 K and 15 K), sample currents ($10$
$\mathrm{\mu A}\sim 300$ $\mathrm{\mu A}$ (and therefore different
electric fields), and sample surface orientations for the two
different \Phos doping concentrations mentioned above. Due to the
variation in charge carrier concentration as well as Fermi- and
quasi-Fermi energies caused by changing these experimental
parameters, the dynamics of the observed spin--dependent
recombination transitions, and therefore the dynamics of the
observed current signals, changed significantly between
measurements. For all data sets, $\tau_0$ was determined for
transients recorded at magnetic fields corresponding to the two
\Phos hyperfine resonances as well as at the maximum of the
overlapping interface defect signals. The results of this procedure
are displayed in Fig.~\ref{fig:5} in two correlation graphs, for
samples with the two differnet donor concentrations. Each graph
displays a plot of $\tau_0^A$ versus $\tau_0^B$ with $A$ and $B$
corresponding to the \Phos low field and the \Phos highfield
resonances, respectively (represented by the circles) the \Phos low
field and the interface state resonances, respectively (represented
by the squares) and the \Phos high field and the interface state
resonances, respectively (represented by the triangles).

%%%%%%%%%%%%%%%%%%%%%%%%%%%%%%%%%%%%%%%%%%%%%%%%%%%%%%%%%%%%%%%%%%%%%
\subsubsection{Discussion}
The data in Fig.~\ref{fig:5} reveals three observations:

(i) As anticipated, the variation of experimental parameters clearly
varies the dynamics of the observed signals.

(ii) While similar combinations of experimental parameters were used
for the two concentrations, the times $\tau_0$ of the samples with
$\mathrm{[P]}=10^{15}$ $\mathrm{cm^{-3}}$ are generally slower than
the $\tau_0$ obtained for $\mathrm{[P]}=10^{16}$ $\mathrm{cm^{-3}}$.
This observation shall not be discussed quantitatively in the
following, however, it is conceivable that as an increase of the
\Phos doping concentration leads to a smaller main--pair distance
(note the definition of main--pair distance given in
section~\ref{maindist}) for \Phos to interface defect pairs as well
as \Phos to \Phos pairs. It should not lead to a change of the
main-pair distance of interface defect pairs. The observed general
decrease of the transition time with an increase of the \Phos
concentration therefore suggests that the observed signals are
predominantly due to transitions involving \Phos and not transitions
between two interface defects. We note that there may be other
reasons for this decrease in transition times with increasing doping
density: one such possibility is that different internal fields may
lead to different energy separations between pair partners with
different intra-pair distance, with a corresponding change in
transition probability.

(iii) The entire set of data reveals a strong correlation of
$\tau_0$ between any combination of signals, whether it is between
the two \Phos hyperfine signals or between the interface signals and
either one of the two \Phos hyperfine signals. The correlation
between the two \Phos hyperfine peaks is expected as the only
difference between \Phos atoms contributing to these two peaks is
the nuclear spin state, which has little influence on electronic
transitions times. The strong correlation between any two of the
\Phos peaks and the interface states is again strong evidence that
the spin--dependent transitions measured involve \Phos--interface
defect pairs in the way sketched in Fig.~\ref{fig:1}(b).

It was not possible to obtain unambiguous correlation plots similar
to those in Fig.~\ref{fig:5} for the two different interface and
near--interface defects discussed above. It is therefore not
possible to verify whether the observed correlation between the
\Phos and the interface defects applies to both interface centers or
only to the one which dominated under the observed conditions.
However, we point out that the strong correlation seen in the plots
of Fig.~\ref{fig:5} is consistent with the assumption that the
dynamics of both defects correlates with the \Phos dynamics and
thus, spin--dependent recombination through both interface defects
in the manner sketched in Fig.~\ref{fig:1}(b) seems possible.

The correlation data of Fig.~\ref{fig:5} raises a question about the
presence of spin--dependent transitions between interface defects.
It is known from previous pEDMR studies of (111) oriented nominally
intrinsic c-Si/\sio interfaces at
X-Band~\cite{Boe:TSF,Frie:05,Boe05}, as well as from recent pEDMR
studies at high magnetic fields ($\approx
8.5\mathrm{T}$)~\cite{mcc:prb:2008}, that spin--dependent
transitions involving only interface defects but not \Phos are
possible and that these transitions can be due to spin--dependent
resonant tunnelling between two interface states in sufficient
spatial and energetic proximity. We conclude from the data seen in
Fig.~\ref{fig:5} that, whilst such interface defect-only processes
may or may not have been present at the investigated c-Si:P/\sio
interfaces, they did not dominate the observed spin--dependent
rates, consistent with the very weak signal strength of previous
pEDMR measurements on intrinsic c-Si/\sio interfaces conducted at
X-Band~\cite{Boe05}. The interface defect signals are weak in spite
of the presence of a significantly higher interface defect density
compared to the areal density of \Phos close to the interface. This
may be explained by considering the signal from interface defect
pairs, which have almost identical g-factors and therefore stronger
coupling, leading to lower pEDMR signals than those seen from
\Phos--defect pairs which have quite different g-factors and weaker,
but still finite, coupling~\cite{gliesche-2008}, and therefore
dominate the signal.

%%%%%%%%%%%%%%%%%%%%%%%%%%%%%%%%%%%%%%%%%%%%%%%%%%%%%%%%%%%%%%%%%%%%%
\subsection{Electrical detection of spin coherence}
%%%%%%%%%%%%%%%%%%%%%%%%%%%%%%%%%%%%%%%%%%%%%%%%%%%%%%%%%%%%%%%%%%%%%

\subsubsection{Experimental data}
\begin{figure}[b]
\centering\includegraphics[width=\columnwidth]{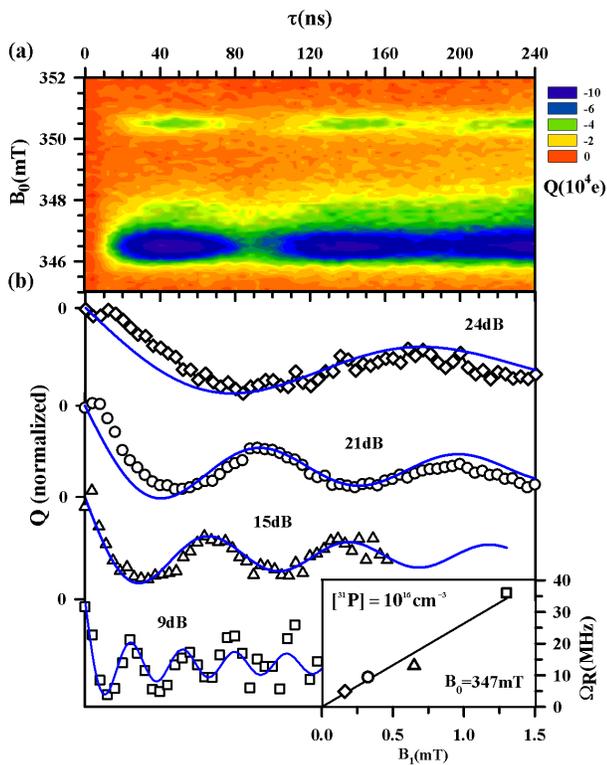}
\caption{(color online) \label{fig:3}(a) Plot of the measured
integrated charge $Q(\tau,B_0)$ as defined by Equ.~\ref{eq1} as a
function of the magnetic field $B_0$ and the length $\tau$ of an
applied microwave pulse with frequency $f = 9.7475$ GHz and power
$P = 8$ W. (b) The symbols represent a plot of the measured charge
$Q(\tau)$ for four different microwave powers at $B_0 = 347$ mT.
Note that for the latter case, the pulse length was recorded up to
$\tau = 120$ ns only as the measurement was limited by signal
perturbation due to the pulse induced microwave current artifacts.
The blue lines represent fits of an integrated Bessel function to
the experimental data. The Rabi--nutation frequencies obtained
from these fits are displayed in the inset as a function of the
applied $B_1$ field. The fit of a linear function through the
origin (black line) shows good agreement.}
\end{figure}

The ability to perform spin--Rabi nutation is a crucial prerequisite
for the coherence time measurements using echo pulse sequences as it
reveals the resonantly induced Rabi frequency for a given set of
experimental conditions (e.g. for the applied microwave power) . The
Rabi frequency is needed to determine the pulse lengths required to
obtain the correct nutation angles during the pulse sequences. We
performed transient nutation style experiments for the electrical
detection of spin--Rabi nutation. For these measurements, the
photocurrent change $\Delta I$ was integrated between two
appropriately chosen integration times $t_1$ and $t_2$ after the
microwave pulse so we obtain a charge
\begin{equation}
Q=\int_{t_1}^{t_2}\Delta I(t)dt \label{eq1}
\end{equation}
that is proportional to the number of spin--dependent transitions
induced by the resonant excitation~\cite{gliesche-2008}. As $Q$ is
proportional to the projection of the resonantly prepared coherent
spin state $|\psi(\tau)\rangle$ at the end of the excitation pulse
of length $\tau$ onto the singlet state
$|S\rangle$~\cite{Boe03b,stegner06}, the measurement of
$Q(\tau)\propto |\langle S|\psi(\tau)\rangle|^2$ reveals the
propagation of the singlet content of $|\psi(\tau)\rangle$ during
the pulse in a transient nutation--style experiment.
Figure~\ref{fig:3}(a) displays the measurement of $Q(\tau,B_0)$ as
function of the pulse length $\tau$ and the magnetic field $B_0$ for
a c-Si sample with $\mathrm{[P]}=10^{16}$ $\mathrm{cm^{-3}}$, a
temperature $T=5$ K, $\theta = 90^\circ$, and integration times
$t_1=6$ $\mu$s and $t_2=16$ $\mu$s.

%figure 5

%%%%%%%%%%%%%%%%%%%%%%%%%%%%%%%%%%%%%%%%%%%%%%%%%%%%%%%%%%%%%%%%%%%%%

\subsubsection{Discussion}
The data set displayed in Fig.~\ref{fig:3} shows that $Q(\tau)$ has
an oscillating behavior around the same magnetic fields which
produced local current response maxima in the data set displayed in
fig.~\ref{fig:2}. The oscillatory dependence on $\tau$ is due to the
dephasing spin--Rabi nutations as demonstrated, for the magnetic
field $B_0=348$ mT, by the data sets displayed in
Fig.~\ref{fig:3}(b): The three plots show $Q(\tau)$ for three
different microwave powers ($B_1$ field strengths). We
anticipate~\cite{Boe03b} the spin--Rabi nutation signal of an
inhomogeneously broadened spin ensemble to follow the integral of a
first kind Bessel function
\begin{equation}
Q(\tau)\propto\int_0^{\gamma B_1\tau}
J_0(2x)dx=\frac{1}{\pi}\int_{-\infty}^\infty
\frac{\sin^2\left(\gamma B_1\tau\sqrt{1+x^2}\right)}{1+x^2}dx
\label{eq2}
\end{equation}
in which $\gamma\approx 2.8$ MHz is the gyromagnetic ratio. A fit of
the three data sets in Fig.~\ref{fig:3}(b) with Equ.~\ref{eq2} shows
a good agreement and it reveals values for the nutation frequencies
$\Omega_R = \gamma B_1$ for the three different powers. A plot of
$\Omega_R$ versus the $B_1$ fields obtained from the relative
microwave attenuation used for the three measurements shows that the
expected linearity of the spin--Rabi nutation is given. The data
shown in Fig.~\ref{fig:3} is thus shown to be due to the spin--Rabi
nutation of \Phos donor and interface electron spins, confirming the
previously reported observation made for c-Si(100):P/\sio interface
with higher \Phos concentration. Note that with the data obtained
from the transient nutation measurement, it is possible to determine
the length of $\pi$- and $\frac{\pi}{2}$--pulses as needed for the
$T_2$ measurements in the following. The latter will be necessary
since the decay of the Rabi nutation as displayed in
Fig.~\ref{fig:3} is not a measure for spin--coherence. The agreement
of the nutation data with the integrated Bessel function is
indicative of coherent dephasing, not coherence decay, being the
dominant source of the observed nutation decay. This assumption is
confirmed by the electrically and pEPR detected echo data discussed
below which shows that the real $T_2$ spin coherence time of the
\Phos donor electrons is distinctively longer than the decay of the
nutation signal.

%%%%%%%%%%%%%%%%%%%%%%%%%%%%%%%%%%%%%%%%%%%%%%%%%%%%%%%%%%%%%%%%%%%%%
\subsection{Comparison of the coherence time $T_2$ of \Phos donor
electrons at the c-Si:P/\sio interface and in the c-Si:P bulk}
%%%%%%%%%%%%%%%%%%%%%%%%%%%%%%%%%%%%%%%%%%%%%%%%%%%%%%%%%%%%%%%%%%%%%

\subsubsection{PEDMR and pEPR detected spin--echoes}

The data presented in sections~\ref{idspdeptrans} and
\ref{idphindeftrans} showed that the spin--dependent current
observed at c-Si:P/\sio interfaces are due to transitions that
take place between \Phos donor electron states and c-Si:P/\sio
interface states in their proximity. A measurement of the \Phos
donor electron spin coherence time (the $T_2$ time) using pEDMR
will therefore reveal only the $T_2$ of \Phos donor electrons in
close proximity of the c-Si:P/\sio interface defects. In contrast,
a $T_2$ measurement using conventional radiation detected pEPR
will reveal the coherence time of \Phos donors in the bulk as the
contribution of near--surface \Phos atoms in the pEPR signal will
be negligible in comparison to the magnitude of the bulk \Phos
signal. Thus, the comparison of $T_2$ times measured by pEDMR and
pEPR allows the influence of interface defects on the \Phos donor
electron spin coherence times to be studied.

\begin{figure}[b]
\centering\includegraphics[width=\columnwidth]{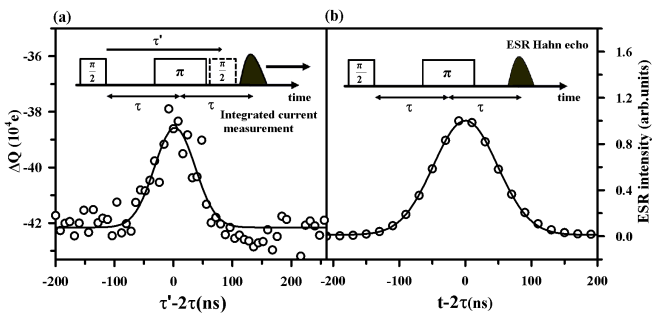}
\caption{Comparison of (a) an electrically detected spin--echo
with $\tau = 300$ ns and (b) a conventional radiation detected
Hahn--echo with $\tau = 10$ $\mu$s. The data sets were recorded on
the same sample, under identical sample conditions (T$ = 10$ K,
$I=250$ $\mu$A, $\theta = 90^\circ$) during the same experimental
run. Both data sets were fit with Gaussian functions (solid line).
Insets of the two plots show sketched timelines of the used pulse
sequences.} \label{fig:6}
\end{figure}

The pEPR and pEDMR experiments require different approaches to the
measurement of $T_2$ times: With pEPR, the $T_2$ times of
paramagnetic centers can be determined most easily using a Hahn-echo
pulse sequence. Hahn--echoes are a temporary rephasing of a
spin--ensemble due to a pulse sequence consisting of an initial
$\pi/2$--pulse which turns the ensemble polarization into the plane
perpendicular to the $B_0$ field and a subsequent $\pi$--pulse which
initiates the phase reversal~\cite{schweiger}. As illustrated in the
inset sketch of Fig.~\ref{fig:6}(b), when the echo pulse sequence
consists of the $\frac{\pi}{2}-\pi$ pulses with a duration $\tau$
between the pulses, a Hahn--echo can be observed at a time $\tau$
after the second pulse which is the time $2\tau$ after he first
pulse. When a Hahn-echo is observed, $T_2$-times can be measured by
determining the decay of the Hahn--echo as a function of twice the
pulse separation time $2\tau$. Figure~\ref{fig:6}(b) shows a data
set for a Hahn--echo transient obtained from a \Phos doped c-Si
sample with $\mathrm{[^{31}P]}=10^{16}$ $\mathrm{cm^{-3}}$ at a
temperature $T = 10$ K and with $\theta = 90^\circ$. One can clearly
see a local maximum of the transient microwave signal at a time
$t=2\tau$. The data set was well fit by a Gaussian function which
was used in order to determine the integrated intensity of the echo.

%figure 6

In contrast to pEPR measurements, pEDMR does not allow direct
observation of Hahn--spin echoes through real time transient
measurements as the integrated sample current $Q(\tau)$ always
represents a projection of the spin--state at the end of the pulse
sequence onto a singlet state. Therefore, the $T_2$ time
measurement using pEDMR requires the utilization of a modified
Hahn-echo pulse sequence that is illustrated in the inset of
Fig.~\ref{fig:6}(a)~\cite{Huebl2008}. In order to resolve the
dynamics of the spin--ensemble during and after the
$\frac{\pi}{2}-\pi$ pulse sequence, a third pulse with length
$\frac{\pi}{2}$ is applied at a time $\tau'$ after the begin of a
conventional Hahn-echo pulse sequence. The third pulse projects
the spin--ensemble at the time $\tau'$ onto the $\hat
z$--direction which in turn determines the singlet content of the
\Phos--interface defect pair. The charge $Q$ integrated following
this pulse therefore represents the polarization of the
spin--ensemble along the $\hat x$- axis [$\hat y$-axis] at the
time $\tau'$ (assuming the $B_1$ field is directed along the $\hat
y$- axis [$\hat x$- axis]). The measurement of the entire echo
transient using the pEDMR detection requires repetition of the
echo sequence for various $\tau'$: the third pulse (also called
the detection pulse) is swept through the time range which covers
the Hahn--echo maximum as well as the echo baseline. While this
procedure makes the time needed for $T_2$ measurements
significantly longer, it allows the measurement of a $T_2$ using
pEDMR. The plot displayed in Fig.~\ref{fig:6}(a) shows an
electrically detected spin--echo measured on the same sample and
under identical conditions, recorded during the same experimental
run as the measurements shown in Fig.~\ref{fig:6}(b). The data
clearly shows the echo whose fit with a Gaussian function is
displayed by the solid line. The comparison of the two echo
functions shows that the electrically detected echo is more narrow
than the radiation detected indicating that it is due to a more
heterogeneous spin ensemble.

The use of two different measurement techniques (pEDMR, pEPR)
raises the question of whether both methods probe the same
observable, namely the $T_2$ of \Phos impurity atoms. There has
recently been a comparative study of $T_2$ times confirming this
identity using a pEDMR detected spin--dependent bulk
process~\cite{Morley2008a} (a spin-trap process of \Phos in c-Si
that becomes relevant at high magnetic fields) which showed that
both pEDMR and pEPR measured $T_2$ times reveal an excellent
agreement. Thus, in the following, systematic measurements
comparing pEPR detected $T_2$-times of \Phos bulk impurities and
pEDMR detected $T_2$-times of \Phos interface impurities are
presented. These measurements are made for three reasons: (i) To
extend the previous observation of electrically detected
Hahn-echoes~\cite{Huebl2008} to c-Si(111) surfaces. (ii) To
measure the temperature dependence of the $T_2$-times. (iii) To
obtain comparable measurements with both pESR and pEDMR under
identical conditions.

%%%%%%%%%%%%%%%%%%%%%%%%%%%%%%%%%%%%%%%%%%%%%%%%%%%%%%%%%%%%%%%%%%%%%
\subsubsection{Measurement of spin--echo decays}
\begin{figure}[t]
\centering\includegraphics[width=0.9\columnwidth]{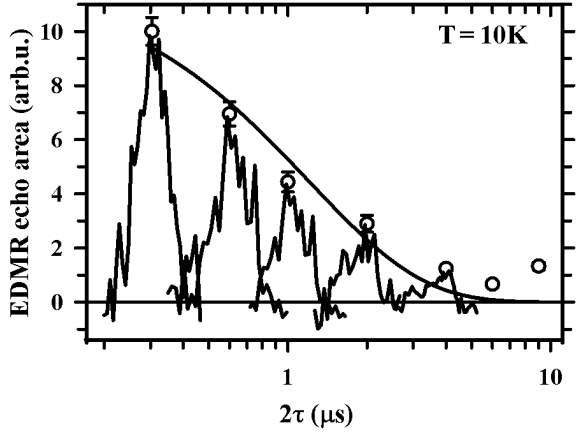}
\caption{Plot of integrated intensities of the electrically
detected echoes as a function of $2\tau$ recorded with pEDMR on a
c-Si:P/\sio sample with $\mathrm{[P]}=10^{16}$ $\mathrm{cm^{-3}}$.
The solid line represents a fit with a modified multiexponential
decay function (see text), the small inset plots represent plots
of echo data sets.
 } \label{fig:7}
\end{figure}

The spin--echo effects shown in Fig.~\ref{fig:6} are imprints of
coherent spin motion on currents or radiation intensities. When a
spin ensemble looses coherence during a $\frac{\pi}{2}-\pi$
sequence, the intensity of the spin--echo following this sequence
decays. Quantifying the decay of Hahn--echoes is a direct measure of
the coherence time $T_2$~\cite{schweiger}. Figure~\ref{fig:7}
displays a set of electrically detected Hahn echoes as well as a
plot of the integrated echo intensities obtained from them, recorded
on a sample with $\mathrm{[P]} = 10^{16}$ $\mathrm{cm}^{-3}$ at
magnetic fields in resonance with the high field \Phos peak at
$T=10$ K and $\theta = 90^\circ$, and with a sample current of
$I=250$ $\mu$A. One can clearly see the gradual decay of the echo
intensity with increasing pulse separation time $\tau$. The solid
line of the plot in Fig.~\ref{fig:7} displays a fit of the intensity
data with a modified exponential function
\begin{equation}
I(2\tau)=e^{-\frac{2\tau}{T_2}-\frac{8\tau^3}{T_S^3}} \label{eq3}
\end{equation}
which contains a contribution due to a single exponential $T_2$
decay as well as the stretched exponential contribution due to the
isotopical influence of the $^{29}$Si which causes spin--diffusion
with time constant $T_S$~\cite{Tyryshkin2003,Morley2008a}. The
plot in Fig.~\ref{fig:7} exhibits a good agreement of the echo
decay with the fit function and thus, by using this method for
both pEDMR as well as pEPR detected echo decay measurements, the
$T_2$ times of \Phos impurities can be determined for the c-Si:P
bulk and the c-Si:P/\sio interface, respectively. Note that all
pESR detected and most pEDMR detected echo measurements were
conducted on the low field peak of the hyperfine split \Phos
resonance. In order to confirm that the spectral proximity of the
\Pb--peak to the low field \Phos resonance under the given
experimental conditions does not distort the electrical $T_2$
measurements, control measurements were carried out on the high
field \Phos peak for $T=5K$. Within the given error margins, the
results of these control measurements (Fig.~\ref{fig:8}, blue
solid diamonds) are in agreement with the measurements obtained
from the low field \Phos resonance.

%figure 7

%%%%%%%%%%%%%%%%%%%%%%%%%%%%%%%%%%%%%%%%%%%%%%%%%%%%%%%%%%%%%%%%%%%%%
\subsubsection{Temperature dependence of $T_2$ times}
\begin{figure}[t]
\centering\includegraphics[width=0.9\columnwidth]{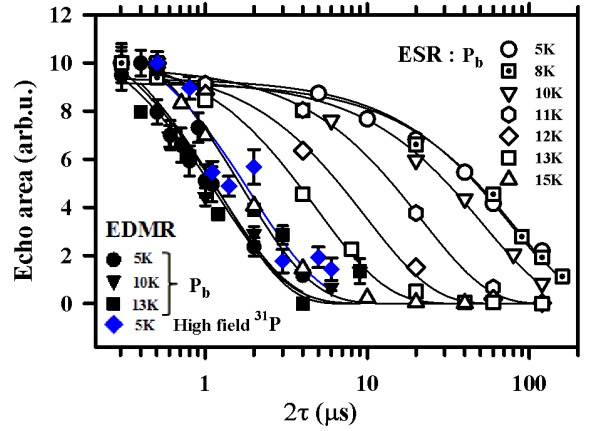}
\caption{(color online) Plot of the normalized integrated echo
intensity of pEPR and pEDMR detected Hahn echoes as a function of
the logarithms of the pulse separation time $\tau$ for various
temperatures. The data sets were fit with the modified exponential
function given by Equ.~\ref{eq3}. All pEDMR and pEPR measurements
were conducted on the same samples during the sample experimental
run.} \label{fig:8}
\end{figure}

The measurement of $T_2$ as described above was repeated on the same
sample with pEPR at temperatures $T=5$ K, 8 K, 10 K, 11 K, 12 K, 13
K, 15 K and with pEDMR at temperatures of $T=5$ K, 10 K, 13 K. The
integrated echo intensities of these measurements are plotted in
Fig.~\ref{fig:8} along with the results of their fit with
Eqn.~\ref{eq3}. The comparison of the pEDMR and the pEPR data sets
shows that while there is a strong temperature dependence of the
echo decay for the pEPR data, the pEDMR decay is faster and, within
the range $T=5$ K to 13 K, nearly constant. In order to analyze
these observations quantitatively, the fit results for $T_2$ are
plotted for both the pEDMR and the pEPR measurements in an Arhenius
plot in Fig.~\ref{fig:9}.

The coherence time of the bulk \Phos donors, $T_2$, determined via
pEPR, is well fit with a function of the form:
\begin{equation}
\frac{1}{T_2}=Ae^{\frac{-\Delta E}{k_B T}}+\frac{1}{T_0}
\label{eq4}
\end{equation}
where $\Delta E$ is an activation energy, and $1/T_0$ a constant
relaxation rate independent of temperature. The best fit occurs with
$\Delta E = 9.1\pm0.5$ meV and $1/T_0 = 10\pm1$ kHz.

%figure 8

%%%%%%%%%%%%%%%%%%%%%%%%%%%%%%%%%%%%%%%%%%%%%%%%%%%%%%%%%%%%%%%%%%%%%
\subsubsection{Discussion}
\label{disT2} 
\begin{figure}[t]
\centering\includegraphics[width=0.85\columnwidth]{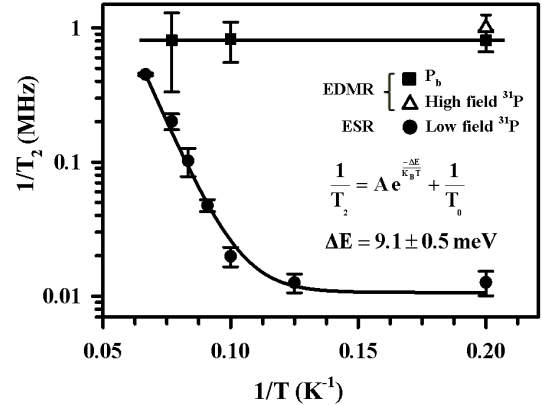}
\caption{Plot of the logarithm of coherence decay rate $T_2^{-1}$
obtained from the data shown in Fig.~\ref{fig:8} as a function of
the inverse temperature $T^{-1}$. The solid lines are fits of the
data. The pEDMR data is fit with a constant function. The pEPR is
fit with a combination of a constant function and a temperature
activated Boltzmann factor.} \label{fig:9}
\end{figure}

As the pEDMR and the pEPR data displayed if
Fig.~\ref{fig:9} was collected on the same sample in the same
experimental run, it is clear that the pEDMR measured $T_2$ times
of \Phos donor spins in close proximity to interface defects are
significantly shorter than the pEPR measured $T_2$ times and,
within the error, independent of the temperature. At the same
time, the pEPR measured $T_2$ of the bulk \Phos donor spins is not
only longer than the pEDMR measured \Phos donor spins, it is also
highly temperature dependent. In the temperature range from $T=5$
K to 15 K, the coherence time changes by about a factor 50,
exhibiting a good agreement with the previously demonstrated
thermal activation~\cite{Tyryshkin2003} of \Phos in a c-Si
environment with a natural abundance of the $^{29}$Si isotope. The
comparison of $T_2$ of \Phos in proximity to interface defects
with those in the bulk therefore shows that interface defects
significantly shorten the donor electron spin coherence time.
Within the given temperature range, $T_2$ appear to be pinned at
$\approx 1.3$ $\mu$s, a value which has been observed previously
for electrically detected \Phos spins in c-Si samples with
different surface orientations, donor concentrations and
experimental conditions~\cite{stegner06, Huebl2008}. The
independence of the donor spin coherence time of near--interface
defect \Phos atoms from experimental conditions, including
temperature, suggests that in contrast to bulk donors, $T_2$ of
the near--surface donors is not determined by spin--spin or
spin--lattice interactions but by transition directly related to
the interface defects in their immediate proximity.

%figure 9

We identify two possible origins for the drastic quenching of the
\Phos donor electron spin $T_2$ time in proximity of interface
defects described above:

(i) The electronic transition between the \Phos donor state and
the \Pb center. This occurs when the donor electron falls into the
doubly occupied interface ground state. The electronic transition
leaves both the \Phos donor and the interface state diamagnetic
since there is no donor spin present after the transition and the
interface state is a doubly occupied singlet state.

(ii) A fast $T_2$ of the interface state which significantly
quenches the $T_2$ relaxation of \Phos donor spins in its
proximity as suggested by de Sousa~\cite{Sousa2007}.

Note that the electronic transition is a limitation to both $T_1$
as well as $T_2$ processes of the spin pairs as it destroys the
pairs and when the electronic transition is the limiting factor
for the $T_2$ times, both, the electrically measured $T_2$ and
$T_1$ times should be equal. In contrast, it is conceivable that
the interface state induced enhancement of the \Phos $T_2$
relaxation keeps the $T_1$ either unchanged or significantly
slower than the $T_2$ time. As a consequence, if the electronic
transition time is longer than the \Pb induced $T_2$ time, the
electrically measured $T_1$ time equals the electronic transition
time and it should therefore be significantly longer than the
electrically measured $T_2$ times. Hence, a study of $T_1$ times
of the near--interface defect \Phos donor electron spins, as
presented in the following, has been conducted.

%%%%%%%%%%%%%%%%%%%%%%%%%%%%%%%%%%%%%%%%%%%%%%%%%%%%%%%%%%%%%%%%%%%%%
\subsection{Comparison of $T_2$ and the longitudinal relaxation time
$T_1$ of \Phos donor electrons at the c-Si:P/\sio interface}
%%%%%%%%%%%%%%%%%%%%%%%%%%%%%%%%%%%%%%%%%%%%%%%%%%%%%%%%%%%%%%%%%%%%%

%%%%%%%%%%%%%%%%%%%%%%%%%%%%%%%%%%%%%%%%%%%%%%%%%%%%%%%%%%%%%%%%%%%%%
\subsubsection{Electrical detection of spin inversion}
\begin{figure}[t]
\centering\includegraphics[width=0.7\columnwidth]{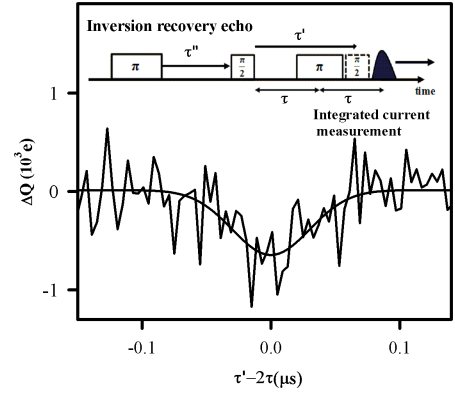}
\caption{Demonstration of an electrically detected spin inversion
recovery experiment. The inset is a sketch of the inversion
recovery pulse sequence which consists of the Hahn--echo sequence
that is preceded by an inversion ($\pi$-) pulse at a time $\tau"$
before the Hahn--echo sequence begins. Similar to the electrically
detected Hahn--echo shown in Fig.~\ref{fig:6}, a projection pulse
is shifted through the sequence during different repetitions of
the experiment. The main plot displays an echo recorded by
plotting $Q$ as a function of the difference $\tau'-2\tau$ between
the projection pulse begin $\tau'$ and the echo maximum at
$2\tau$.} \label{fig:10}
\end{figure}

In order to probe $T_1$ of near--interface \Phos, electrically
detected inversion recovery experiments were carried out for a
temperature of $T=5\mathrm{K}$ where pESR measurements of bulk
\Phos donor spins reveal large differences between $T_1$ and $T_2$
times~\cite{Tyryshkin2003}. The idea behind the inversion recovery
experiment~\cite{schweiger} is to invert a given spin polarization
towards a negative non--equilibrium polarization and to then
observe transiently the gradual return toward the equilibrium due
to $T_1$ processes. Thus, the experiment consists of an initial
inversion pulse, a defined delay time $\tau"$ during which the
spin ensemble undergoes $T_1$ relaxation and a subsequent
polarization measurement which is typically conducted by
utilization of a Hahn--echo sequence with fixed pulse separation
time $\tau$. Figure~\ref{fig:10} displays a sketch of the pulse
sequence used for the electrical inversion recovery measurements
presented in the following. Similar to the Hahn--echo decay
measurements, the Hahn--echo is measured with pEDMR by repetition
of the experiment whilst applying projection pulses which are
gradually swept through the echo sequence. The data in
Fig.~\ref{fig:10} shows an inverted Hahn--echo recorded with a
short $\tau=252\mathrm{\mu s}$ and an equally short delay time
$\tau"=52ns$ producing strong inversion. Instead of the positive
spin echo as detected for a simple Hahn--echo sequence, the sign
of the echo is negative. To the knowledge of the authors, the data
shown in Fig.~\ref{fig:10} is the first demonstration of an
electrically detected spin inversion experiment.

%figure 10

%%%%%%%%%%%%%%%%%%%%%%%%%%%%%%%%%%%%%%%%%%%%%%%%%%%%%%%%%%%%%%%%%%%%%
\subsubsection{Inversion recovery of near--interface defect $^{31}\mathrm{P}$ donor electrons}
\begin{figure}[t]
\centering\includegraphics[width=0.8\columnwidth]{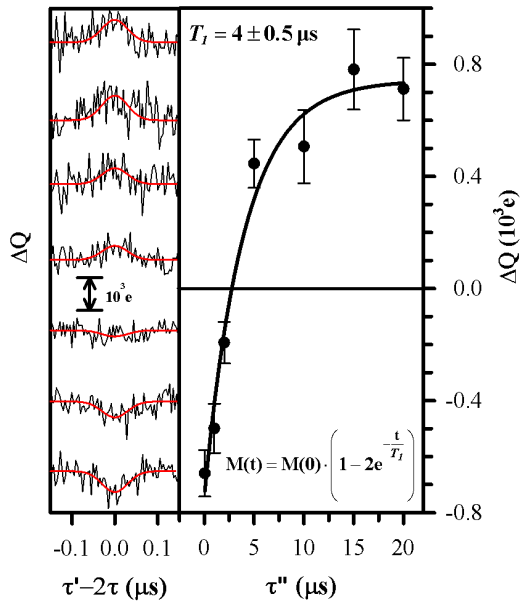}
\caption{(color online) Plot of the integrated echo intensity
detected with the pulse sequence shown in Fig.~\ref{fig:10} as a
function of the inversion recovery time $\tau"$. The solid line
represents a fit of the data with a single exponential function.
Note that the inverted echo for small $\tau"$ changes into a
non--inverted echo with equal magnitude for large $\tau"$. The
inset plots on the left show the raw data of the various echo
measurements as well as fits with Gaussian functions which are the
basis for the integrated echo intensities.} \label{fig:11}
\end{figure}

In order to determine $T_1$, the experiment shown in
Fig.~\ref{fig:10} was repeated for seven values of $\tau"$ in the
range of $52\mathrm{ns}\leq\tau"\leq20\mathrm{\mu s}$. The results
of these experiments are displayed in Fig.~\ref{fig:11}. They show
that the polarization inversion that exists directly after the
inversion pulse ($\tau"$ is very small) exponentially approaches
the steady state polarization with increasing $\tau"$. The
integrated echo amplitudes were obtained from a fit of the echo
data with Gaussian functions. They dependence on $\tau"$ show an
excellent agreement with an exponential decay function with a
negative offset
\begin{equation}
M(\tau")=M(0)\left[1-2e^{-\frac{t}{T_1}}\right] \label{eq5}
\end{equation}.
The time constant $T_1=4.0(5)\mathrm{\mu s}$ obtained from this
fit is more than six orders of magnitude shorter than the
previously investigated bulk $T_1$ times~\cite{Tyryshkin2003}
which shows that the proximity of \Pb centers leads to
dramatically reduced $T_1$ times.

%figure 11

%%%%%%%%%%%%%%%%%%%%%%%%%%%%%%%%%%%%%%%%%%%%%%%%%%%%%%%%%%%%%%%%%%%%%
\subsubsection{Discussion}
The measurement of $T_1$ of \Phos donor electrons in proximity of
interface defects revealed a $T_1$ time which was quenched by
orders of magnitude compared to bulk \Phos: However, in contrast
to bulk pESR measurements, $T_1$ is only about a factor 3 longer
than the electrically detected $T_2$ time at the measured
temperature of $T=5\mathrm{K}$. Following the argumentation given
in section~\ref{disT2}, we learn that since $T_2<T_1$, $T_1$ will
be determined by the electronic transition but not $T_2$ as it is
faster due to the influence of the nearby interface defect. It
shall be emphasized though that the errors of the $T_1$ and $T_2$
measurements have been deduced from the fits of the Hahn--echo and
inversion recovery data with eqs.~\ref{eq3} and \ref{eq5},
respectively which leaves a significant uncertainty regarding the
magnitude of the measured $T_1$ and $T_2$ times. Given the
proximity of the two values within the calculated errors and the
remaining uncertainty about the given errors, it can not be
absolutely excluded that $T_1\approx T_2$. The latter would be
indicative that the electronic transition not only determines
$T_1$ but also $T_2$. Note that in spite of this uncertainty, is
is clear that the electronic transition can not be slower than the
measured $T_1$-time and it can not be faster than the measured
$T_2$-time. Hence it is unambiguously clear that the
\Phos--interface defect process takes place on a time scale in the
lower microsecond range ($1\mathrm{\mu s}-4\mathrm{\mu s}$).

%%%%%%%%%%%%%%%%%%%%%%%%%%%%%%%%%%%%%%%%%%%%%%%%%%%%%%%%%%%%%%%%%%%%%
\section{Conclusions}
The realization that $T_1$ and $T_2$ times of \Phos donor electron
spins near--c-Si:P/\sio interface defects are drastically
shortened to a similar value is proof that the recombination of
the \Phos donor electron into a charged interface ground state is
the lifetime limiting process of the \Phos qubit. This conclusion
is also supported by the independence of the transition time on
the temperature as this suggests that the observed spin relaxation
times are determined by the \Phos--\Pb pair geometry and not
phonon density. This conclusion is also consistent with the
assumption that only pairs with a certain, narrowly distributed
"main" distance contribute to the observed pEDMR signals.

Our results have implications for proposed mechanisms to readout
donor spins using nearby probe spins, particularly interface
defects. Without the ability to vary the coupling between donor
and probe spins, the electronic transition times will be fixed,
limiting $T_2$ and also limiting the readout time. To overcome
this problem, control of the coupling between target and probe
spins will be required. This may be achievable using the Stark
effect as electric fields will have a stronger effect on the
localization of the slightly delocalized \Phos donor wave function
in comparison to the extraordinary strongly localized deep
interface defects. Whether these different electric field
sensitivities have a significant impact on the exchange coupling
between the \Phos donor interface states in close proximity at
sufficiently low fields remains to be proven.

The data presented and discussed above strongly supports the model
for spin--dependent recombination via \Phos donors and interface
defects that was presented by Stegner et al.~\cite{stegner06}.
Whilst this explanation for the observed EDMR signal has become
commonly accepted, and is further supported by the results presented
here, the possibility remains that other processes also contribute
to the resonant changes in current. Given the now significant
evidence supporting the $^{31}$P-\Pb
model~\cite{McCamey2006,stegner06,Huebl2008}, we agree with the
conventional understanding and conclude here that we are indeed
observing spin dependent transitions between \Phos donors located
close to \Pb defects at the Si(111)-\sio interface.

Finally, we note that the effect of spin--dependent transitions in
reducing coherence times may not be restricted to the mechanisms
presented here, but may also be important for other mechanisms, such
as spin dependent scattering of conduction electrons by donor spins
in two dimensional electron gasses
~\cite{Ghosh1992,Lo2007,Beveren2008, Sarovar2008,Sousa2008}.

%%%%%%%%%%%%%%%%%%%%%%%%%%%%%%%%%%%%%%%%%%%%%%%%%%%%%%%%%%%%%%%%%%%%%
\section{Summary}
In summary, we have reported on the investigation of
spin--dependent processes at the c-Si:P/\sio interface using
pEDMR, and shown that spin--dependent \Phos to interface defect
recombination takes place at the c-Si (111) surface in a similar
way as previously described processes for c-Si(100) surfaces. The
imprints of spin--dependent recombination on interface currents
reveal EPR resonances of the hyperfine split \Phos resonance and
\Pb defects and at least one other defect, assigned here to the
unrelaxed $E'$ defect in the \sio matrix, previously unobserved in
\Phos doped samples. The correlation measurements of the dynamics
of these pEDMR detected signals strongly support the model that
recombination transitions between \Phos and the interface defect
states occur. By electrical detection of spin--echoes, a
measurement of both the coherence time $T_2$ as well as the
longitudinal relaxation time $T_1$ of \Phos donor spins in
proximity to interface defects was possible. These measurements
revealed that $T_2\approx 1.3$ $\mu$s independent of the applied
temperature range for $5\mathrm{K}\leq T \leq 13\mathrm{K}$. The
longitudinal relaxation time $T_1\approx 4\mathrm{\mu s}$ at
$T=5\mathrm{K}$. The observations contrast the pESR measured $T_1$
and $T_2$ of bulk \Phos which are significantly longer and
strongly temperature dependent. Our observation show that the
proximity of interface defects to \Phos donors introduces
spin--dependent recombination and that these electronics
transitions significantly limit both, the phase coherence time
$T_2$ as well as the longitudinal relaxation time $T_1$ of donors
near interface defects. The implications of these findings for
possible applications of the \Phos--interface defect transition as
spin \Phos readout for proposed potential spin electronics or
quantum information applications have been discussed.

\newpage
%\begin{acknowledgments}
%\end{acknowledgments}

%\bibliographystyle{prsty}

\end{document}